\documentclass[twocolumn,aps,superscriptaddress,showpacs]{revtex4}
\usepackage{amssymb}
\usepackage{amsmath,bm}
\usepackage{graphicx}
\usepackage[normalem]{ulem}
\usepackage[dvips]{color}

\renewcommand\sout{\bgroup \color{red} \ULdepth=-.5ex \ULset}

\begin{document}

\title{Constraining the symmetry energy at subsaturation densities
using isotope binding energy difference and neutron skin thickness}
\author{Zhen Zhang}
\affiliation{Department of Physics and Astronomy and Shanghai Key Laboratory for
Particle Physics and Cosmology, Shanghai Jiao Tong University, Shanghai 200240, China}
\author{Lie-Wen Chen\footnote{%
Corresponding author: lwchen$@$sjtu.edu.cn}}
\affiliation{Department of Physics and Astronomy and Shanghai Key Laboratory for
Particle Physics and Cosmology, Shanghai Jiao Tong University, Shanghai 200240, China}
\affiliation{Center of Theoretical Nuclear Physics, National Laboratory of Heavy Ion
Accelerator, Lanzhou 730000, China}
\date{\today}

\begin{abstract}
We show that the neutron skin thickness $\Delta r_{np}$ of heavy nuclei is
uniquely fixed by the symmetry energy density slope $L({\rho })$ at a
subsaturation cross density $\rho_c \approx 0.11$ fm$^{-3}$ rather than at
saturation density $\rho_0$, while the binding energy difference $\Delta E$
between a heavy isotope pair is essentially determined by the magnitude of
the symmetry energy $E_{\text{sym}}({\rho })$ at the same $\rho_c$.
Furthermore, we find a value of $L({\rho_c })$ leads to a negative
$E_{\text{sym}}({\rho _{0}})$-$L({\rho _{0}})$ correlation while a value of
$E_{\text{sym}}({\rho _{c})}$ leads to a positive one. Using data on
$\Delta r_{np}$ of Sn isotopes and $\Delta E$ of a number of heavy
isotope pairs, we obtain simultaneously $E_{\text{sym}}({\rho _{c})}=26.65\pm0.20$ MeV
and $L({\rho_c })= 46.0\pm4.5$ MeV at $95\%$ confidence level, whose
extrapolation gives $E_{\text{sym}}({\rho _{0}})=32.3\pm1.0$ MeV and
$L({\rho _{0}})=45.2\pm10.0$ MeV. The implication of these new constraints
on the $\Delta r_{np}$ of $^{208}$Pb and the core-crust transition
density in neutron stars is discussed.
\end{abstract}

\maketitle

\section{Introduction}

The determination of density dependence of the symmetry energy
$E_{\text{sym}}(\rho )$, which characterizes the isospin dependent part of
the equation of state (EOS) of asymmetric nuclear matter, is of fundamental
importance due to its multifaceted roles in nuclear physics and
astrophysics~\cite{Lat04,Ste05,Bar05,LCK08} as well as some issues of new
physics beyond the standard model~\cite{Hor01b,Sil05,Wen09}.
Due to the particularity of nuclear saturation density $\rho_0$
($\sim 0.16$ fm$^{-3}$), a lot of works have been devoted to
constraining quantitatively the magnitude and density slope of the symmetry
energy at $\rho_0$, i.e., $E_{\text{sym}}(\rho_0)$ and $L(\rho_0)$, by analyzing
terrestrial nuclear experiments and astrophysical observations. Although
significant progress has been made during the last decade, large uncertainties on
$E_{\text{sym}}(\rho_0)$ and $L(\rho_0)$ still exist (See, e.g.,
Refs.~\cite{Bar05,Ste05,LCK08,Tsa12,Lat12,ChenLW12,LiBA12}).
For example, while the value of $E_{\text{sym}}(\rho_0)$ is determined to be
around $30\pm 4$ MeV, the extracted $L(\rho_0)$ varies drastically
from about $20$ to $115$ MeV, depending on the observables and analysis
methods. To better understand the model dependence of the constraints and
reduce the uncertainties is thus of critical importance and remains a big
challenge in the community. In this work, we show the isotope binding energy
difference and neutron skin thickness of heavy nuclei can be used to stringently
constrain the subsaturation density behavior of the symmetry energy.

The neutron skin thickness
$\Delta r_{np}=\langle r_{n}^{2}\rangle ^{1/2}-\langle r_{p}^{2}\rangle ^{1/2} $
of heavy nuclei, i.e., the difference of the neutron and proton
rms radii, has been shown to be a good probe of $E_{\text{sym}}(\rho )$~\cite%
{Bro00,Hor01,Fur02,Yos04,Che05b,Tod05,Cen09,Rei10,Roc11,Agr12,Oya03,Dan03},
and this provides a strong motivation for the Lead Radius Experiment (PREX)
being performed at the Jefferson Laboratory to determine the
$\langle r_{n}^{2}\rangle ^{1/2}$ of $^{208}$Pb to about $1\%$ accuracy by
measuring the parity-violating electroweak asymmetry in the elastic scattering
of polarized electrons from $^{208}$Pb~\cite{PREX,Abr12,Hor12}. Physically, the
$\Delta r_{np}$ depends on the pressure of neutron-rich matter in nuclei
which will balance against the pressure due to nuclear surface
tension~\cite{Hor01}. Since the pressure of neutron-rich matter is essentially
controlled by the density dependence of $E_{\text{sym}}(\rho )$ and the
characteristic (average) density in finite nuclei is less than $\rho_0$ (See,
e.g., Ref.~\cite{Kha12}), one expects that the $\Delta r_{np}$ should depend on
the subsaturation density behaviors of the $E_{\text{sym}}(\rho )$~\cite{Ste05}.
Brown and Typel~\cite{Bro00} noted firstly that the $\Delta r_{np}$ of heavy nuclei
from model calculations is linearly correlated with the pressure of pure
neutron matter at a subsaturation density of $0.1$ fm$^{-3}$. The linear
correlation of the $\Delta r_{np}$ with both $E_{\text{sym}}(\rho_0)$ and
$L(\rho_0)$ has also been observed in mean-field
calculations~\cite{Bro00,Hor01,Fur02,Yos04,Che05b,Tod05,Cen09,Rei10,Roc11,Agr12}
using many existing nuclear effective interactions.

Recently, a remarkable negative correlation between $E_{\text{sym}}(\rho_0)$
and $L(\rho_0)$ has been obtained by analyzing existing data on
$\Delta r_{np}$ of Sn isotopes~\cite{Che10}, showing a striking contrast to
other constraints that essentially give a positive
$E_{\text{sym}}(\rho_0)$-$L(\rho_0)$ correlation (See, e.g.,
Refs.~\cite{Lat12,ChenLW12,LiBA12}). A negative correlation between
$E_{\text{sym}}(\rho_0)$ and $L(\rho_0)$ has also been observed for a fixed
value of $\Delta r_{np}$ in $^{208}$Pb~\cite{Ste09}. It is thus of great
interest to understand the physics behind this negative
$E_{\text{sym}}(\rho_0)$-$L(\rho_0)$ correlation. Using a recently developed
correlation analysis method~\cite{Che10}, we show here that the $\Delta r_{np}$
of heavy nuclei is uniquely fixed by the $L({\rho })$ at a subsaturation
cross density $\rho_c \approx 0.11$ fm$^{-3}$, which naturally leads to a
negative $E_{\text{sym}}({\rho _{0}})$-$L({\rho _{0}})$ correlation.
Furthermore, we demonstrate that the binding energy difference
between a heavy isotope pair is essentially determined by the magnitude of
$E_{\text{sym}}({\rho })$ at the same $\rho_c$. For the first time, we obtain
simultaneously in the present work stringent constraints on both the magnitude
and density slope of the $E_{\text{sym}}({\rho })$ at $\rho_c \approx 0.11$
fm$^{-3}$ by analyzing data of the isotope binding energy difference for heavy
nuclei and $\Delta r_{np}$ of Sn isotopes, which has important implications on
the values of $E_{\text{sym}}(\rho_0)$ and $L(\rho_0)$, the $\Delta r_{np}$ of
$^{208}$Pb, and the core-crust transition density $\rho_t$ of neutron stars.

\section{Model and method}

The EOS of asymmetric nuclear matter at baryon density $\rho $ and
isospin asymmetry $\delta=(\rho _{n}-\rho _{p})/(\rho _{p}+\rho _{n})$,
given by its binding energy per nucleon, can be expanded to $2$nd-order
in $\delta $ as
\begin{equation}
E(\rho ,\delta )=E_{0}(\rho )+E_{\mathrm{sym}}(\rho )\delta ^{2}+O(\delta
^{4}),  \label{EOSANM}
\end{equation}%
where $E_{0}(\rho )=E(\rho ,\delta =0)$ is the binding energy per nucleon in
symmetric nuclear matter, and the nuclear symmetry energy is expressed as
\begin{equation}
E_{\mathrm{sym}}(\rho )=\frac{1}{2!}\frac{\partial ^{2}E(\rho ,\delta )}{%
\partial \delta ^{2}}|_{\delta =0}.  \label{Esym}
\end{equation}%
Around a reference density $\rho _{r}$, the $E_{\mathrm{sym}}(\rho )$ can be
characterized by using the value of $E_{\text{sym}}({\rho _{r}})$
and the density slope parameter $L(\rho _{r})=3{\rho _{r}}\frac{\partial E_{\mathrm{sym}}(\rho )}{\partial
\rho }|_{\rho ={\rho _{r}}}$, i.e.,
\begin{equation}
E_{\text{sym}}(\rho )=E_{\text{sym}}({\rho _{r}})+L(\rho _{r})\chi_r+O(\chi_r^2),
\end{equation}
with $\chi_r=(\rho -{\rho _{r}})/3\rho _{r}$.

In the present work, we use the Skyrme-Hartree-Fock (SHF) approach with the
so-called standard form of Skyrme force (see, e.g., Ref.~\cite{Cha97})
which includes $10$ parameters, i.e., the $9$ Skyrme force parameters
$\sigma $, $t_{0}-t_{3}$, $x_{0}-x_{3}$, and the spin-orbit coupling constant
$W_{0}$. This standard SHF approach has been shown to be very successful in
describing the structure of finite nuclei, especially global properties such
as binding energies and charge radii~\cite{Cha97,Fri86,Klu09}. Instead
of using directly the $9$ Skyrme force parameters, we can express them
explicitly in terms of $9$ macroscopic quantities, i.e., $\rho _{0}$,
$E_{0}(\rho_{0})$, the incompressibility $K_{0}$, the isoscalar effective
mass $m_{s,0}^{\ast }$, the isovector effective mass $m_{v,0}^{\ast }$,
$E_{\text{sym}}({\rho _{r}})$, $L({\rho _{r}})$, $G_{S}$, and $G_{V}$.
The $G_{S}$ and $G_{V}$ are respectively the gradient and symmetry-gradient
coefficients in the interaction part of the binding energies for finite
nuclei defined as
\begin{equation}
E_{\mathrm{grad}}=G_{S}(\nabla \rho )^{2}/(2{\rho )}-G_{V}\left[ \nabla
(\rho _{n}-\rho _{p})\right] ^{2}/(2{\rho )}.
\end{equation}
Then, by varying individually these macroscopic quantities within their
known ranges, we can examine more transparently the correlation of properties
of finite nuclei with each individual macroscopic quantity. Recently, this
correlation analysis method has been successfully applied to study the
neutron skin~\cite{Che10} and giant monopole resonance of finite
nuclei~\cite{Che12}, the higher order bulk
characteristic parameters of asymmetric nuclear matter~\cite{Che11a}, and
the relationship between the nuclear matter symmetry energy and the symmetry
energy coefficient in the mass formula~\cite{Che11}, where the reference
density $\rho _{r}$ has been set to be $\rho _{0}$.

\section{Results and discussions}

\begin{figure}[tbp]
\includegraphics[scale=0.75]{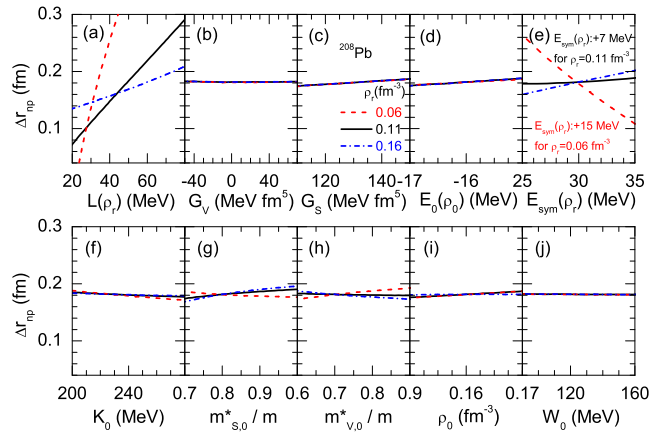}
\caption{(Color online) The $\Delta r_{np}$ of $^{208}$Pb from SHF with MSL0
by varying individually $L(\rho_r)$ (a), $G_{V}$ (b), $G_{S}$ (c),
$E_{0}(\rho _{0})$ (d), $E_{\text{sym}}(\rho _{r})$ (e), $K_{0}$ (f),
$m_{s,0}^{\ast }$ (g), $m_{v,0}^{\ast }$ (h), $\rho _{0}$ (i), and $W_{0}$ (j)
for $\rho_r=0.06$, $0.11$, and $0.16{~\text{fm}}^{-3}$. The $E_{\text{sym}}(\rho _{r})$
is shifted by adding $15$ and $7$ MeV for $\rho _{r}=0.06$
and $0.11{~\text{fm}}^{-3}$, respectively.}
\label{NskinPb208}
\end{figure}
To examine the correlation of the $\Delta r_{np}$ of heavy nuclei with each
macroscopic quantity, especially on $E_{\text{sym}}({\rho _{r}})$
and $L(\rho_r)$, we show in Fig.~\ref{NskinPb208} the $\Delta r_{np}$ of $^{208}$Pb
from SHF with the Skyrme force MSL0~\cite{Che10} by varying individually $L(\rho_{r})$,
$G_{V}$, $G_{S}$, $E_{0}(\protect\rho _{0})$, $E_{\text{sym}}(\rho _{r})$,
$K_{0}$, $m_{s,0}^{\ast }$, $m_{v,0}^{\ast }$, $\rho _{0}$, and $W_{0}$ within
their empirical uncertain ranges, namely, varying one quantity at a time while
keeping all others at their default values in MSL0, for three values
of $\rho_r$, i.e., $\rho_r=0.06$, $0.11$, and $0.16~{\text{fm}}^{-3}$. It is
seen from Fig.~\ref{NskinPb208} that the $\Delta r_{np}$ of $^{208}$Pb exhibits
a strong correlation with $L(\rho_{r})$ and $E_{\text{sym}}(\rho _{r})$ while
much weak correlation with other macroscopic quantities. In particular,
while the $\Delta r_{np}$ always increases with $L(\rho_{r})$, it can increase
or decrease with $E_{\text{sym}}(\rho _{r})$, depending on the value of
$\rho_r$. Most interestingly, one can see that, for
$\rho_r=\rho_c \approx 0.11~{\text{fm}}^{-3}$, the $\Delta r_{np}$ becomes
essentially independent of $E_{\text{sym}}(\rho _{r})$ and only sensitive to
the value of $L(\rho_{r})$. These features imply that the $L({\rho_c})$ is a
unique quantity to determine the $\Delta r_{np}$ of heavy nuclei, and the
experimental data on $\Delta r_{np}$ of heavy nuclei can put
strong limit on the value of $L({\rho_c})$.

\begin{figure}[tbp]
\includegraphics[scale=0.75]{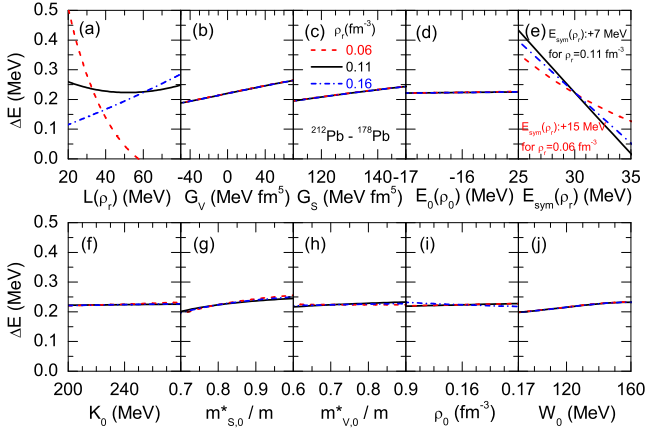}
\caption{(Color online) Same as Fig.~\ref{NskinPb208} but for $\Delta E$
between isotope pair $^{212}$Pb and $^{178}$Pb.}
\label{dEPb}
\end{figure}
In order to determine the magnitude of the symmetry energy at $\rho _{c}$,
i.e., $E_{\text{sym}}(\rho _{c})$, we propose here to use the difference of
binding energy per nucleon between an isotope pair, denoted as $\Delta E$.
Before detailed quantitative calculations, it is instructive to estimate
the $\Delta E$ from the well-known semiempirical nuclear mass formula in
which the binding energy per nucleon for a nucleus with $N$ neutrons and
$Z$ protons ($A=N+Z$) can be approximated by
\begin{eqnarray}
E(N,Z) &=&a_{\text{vol}}+a_{\mathrm{surf}}A^{-1/3}+a_{\text{sym}}(A)\left( \frac{N-Z}{A}\right) ^{2}
\notag \\
&&+a_{\text{Coul}}\frac{Z(Z-1)}{A^{4/3}}+E_{\text{pair}}.  \label{LDM}
\end{eqnarray}%
For heavy spherical even-even nuclei, the $\Delta E$ can then be expressed
approximately as
\begin{eqnarray}
\Delta E &=& E(N+\Delta N,Z) - E(N,Z)
\notag \\
 &\approx& a_{\text{sym}}(A)\frac{4Z(N-Z)}{A^{2}}\times\frac{\Delta N}{A} \notag \\
 &&-a_{\text{Coul}}\frac{4Z(Z-1)}{3A^{4/3}}\times\frac{\Delta N}{A}
 - \frac{a_{\mathrm{surf}}}{3A^{1/3}}\times\frac{\Delta N}{A}, \label{dELDM}
\end{eqnarray}%
if we assume $\Delta N$ is significantly less than $A$ and $N-Z$ and
$a_{\text{sym}}(A+\Delta N)\approx a_{\text{sym}}(A)$. Since the Coulomb
term is relatively well-known and the contribution of the surface term to
$\Delta E$ is generally small compared to that of the symmetry energy
term for heavy neutron-rich nuclei (empirically we have
$a_{\text{sym}}(A)\sim 25$ MeV and $a_{\mathrm{surf}}\sim 18$ MeV), thus
the $\Delta E$ essentially reflects the symmetry energy of finite nuclei,
i.e., $a_{\text{sym}}(A)$. For heavy nuclei, one has the empirical relation of
$a_{\text{sym}}(A)\approx E_{\text{sym}}(\rho _{c})$~\cite{Cen09,Che11}, and
thus expects that the $\Delta E$ for heavy isotope pairs should be a good
probe of $E_{\text{sym}}(\rho _{c})$. To see more quantitative results,
similarly as in Fig.~\ref{NskinPb208}, we show in Fig.~\ref{dEPb} the
$\Delta E$ for isotope pair $^{212}$Pb and $^{178}$Pb from SHF with MSL0 by
varying individually the $9$ macroscopic quantities and $W_{0}$ for
$\rho_r=0.06$, $0.11$ and $0.16~{\text{fm}}^{-3}$. As expected, it is seen
from Fig.~\ref{dEPb} that, for $\rho_{r}=0.11~{\text{fm}}^{-3}$, the
$\Delta E$ exhibits a very strong correlation with $E_{\text{sym}}(\rho _{r})$
while it displays relatively weak dependence on $L(\rho_r)$, $G_{V}$,
$G_{S}$, $m_{s,0}^{\ast }$ and $W_{0}$, and essentially no dependence on other
macroscopic quantities. For $\rho_{r}=0.06$ and $0.16~{\text{fm}}^{-3}$, the
$\Delta E$ also exhibits strong correlation with $L(\rho_{r})$. These results
indicate that the $\Delta E$ for heavy isotope pairs indeed provides a good
probe of $E_{\text{sym}}({\rho _{c}})$. We note here that using other
standard Skyrme forces instead of MSL0 or using other heavy nuclei such as
Sn isotopes leads to similar results as shown in Fig.~\ref{NskinPb208} and
Fig.~\ref{dEPb}. We have also checked with a number of existing standard
Skyrme forces and obtained the similar conclusion~\cite{Zha13}.

\begin{figure}[tbp]
\includegraphics[scale=0.8]{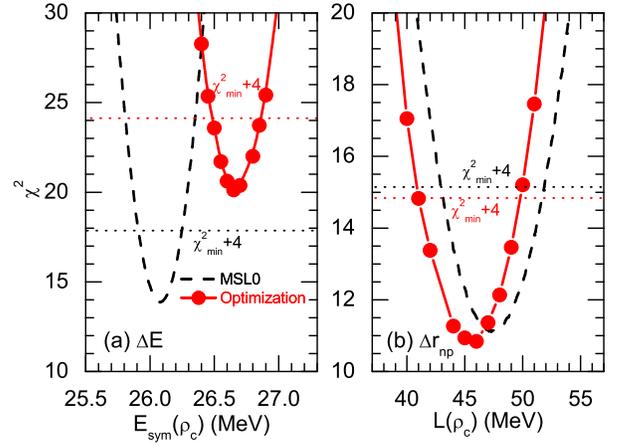}
\caption{(Color online) ${\chi}^2$ as a function of $E_{\text{sym}}(\rho_c)$ (a)
and $L(\rho_c)$ (b). Dashed lines represent results from SHF with MSL0 while the
solid circles are obtained by optimization.}
\label{chsqLcEsymc}
\end{figure}

Experimentally, there are very rich and accurate data on ground state binding
energy of finite nuclei. To constrain $E_{\text{sym}}({\rho _{c}})$ from
$\Delta E$, here we select $19$ heavy isotope pairs which are all spherical
even-even nuclei, namely,
$^{218,206}_{\,~~~~~86}$Rn, $^{216,194}_{\,~~~~~84}$Po, $^{214,178}_{\,~~~~~82}$Pb, $^{212,178}_{\,~~~~~82}$Pb,
$^{210,178}_{\,~~~~~82}$Pb, $^{208,178}_{\,~~~~~82}$Pb, $^{206,178}_{\,~~~~~82}$Pb, $^{206,172}_{\,~~~~~80}$Hg,
$^{136,106}_{\,~~~~~52}$Te, $^{132,100}_{\,~~~~~50}$Sn, $^{132,102}_{\,~~~~~50}$Sn, $^{132,104}_{\,~~~~~50}$Sn,
$^{132,106}_{\,~~~~~50}$Sn, $^{132,110}_{\,~~~~~50}$Sn, $^{132,114}_{\,~~~~~50}$Sn, $^{130,98}_{\,~~~~48}$Cd, $^{124,96}_{\,~~~~46}$Pd, $^{94,84}_{~~~42}$Mo, and $^{94,82}_{~~~40}$Zr. On the other hand, the
$\Delta r_{np}$ of heavy Sn isotopes has been systematically measured using
various methods, and here we use the existing $21$ data on $\Delta r_{np}$ of
Sn isotopes~\cite{Ray79,Kra94,Kra99,Trz01,Kli07,Ter08} to constrain the
$L({\rho _{c}})$. Firstly, with all other parameters fixed at their default
values in MSL0, we calculate the ${\chi}^2$ from the difference between
theoretical and experimental $\Delta E$ ($\Delta r_{np}$) with different
$E_{\text{sym}}({\rho _{c}})$ ($L({\rho}_{c})$), and the results are
shown by dashed lines in Fig.~\ref{chsqLcEsymc}. For the evaluation of
${\chi}^2$, since the experimental precision of the binding energy is much
better than what one can expect from the mean-field description due to the
model limitation, here we adopt the usual strategy (See, e.g.,~\cite{Klu09}),
namely, to assign a theoretical error to the binding energy. In particular,
when we evaluate the ${\chi}^2$, we use the experimental errors for
$\Delta r_{np}$ while assign a theoretical relative error of $23\%$ for
$\Delta E$ so that the minimum value of ${\chi}^2$ is close to the number
of data points to make the ${\chi}^2$-analysis valid~\cite{Bev03}. From the
dashed lines in Fig.~\ref{chsqLcEsymc} obtained with MSL0, one can extract
a value of $E_{\text{sym}}({\rho _{c}})=26.08\pm0.17$ MeV and
$L({\rho}_{c})=47.3\pm4.5$ MeV at $95\%$ confidence level.

The values of $E_{\text{sym}}({\rho _{c}})=26.08\pm0.17$ MeV and
$L({\rho}_{c})=47.3\pm4.5$ MeV are obtained by assuming the other
macroscopic quantities are fixed at their default values in MSL0 when
$E_{\text{sym}}({\rho _{c}})$ or $L({\rho}_{c})$ is varied, and thus the
correlations of $\Delta E$ ($\Delta r_{np}$) with the other macroscopic
quantities are totally neglected. To be more precise, one should consider the
possible variations of the other macroscopic quantities due to the correlations
introduced by fitting the theoretical predictions to some well-known experimental
observables or empirical values. Taking this into account, for a fixed
$E_{\text{sym}}({\rho _{c}})$ ($L({\rho}_{c})$), we optimize all the other
parameters, instead of simply keeping them at their default values in MSL0,
by minimizing the weighted sum of ${\chi}^2$ evaluated from the difference
between SHF prediction and the experimental data for some selected observables
using the simulated annealing method (See, e.g., Ref.~\cite{Agr05}). In the optimization, we
select the binding energy per nucleon $E_{B}$ and charge rms radii $r_c$
of $25$ spherical even-even nuclei, i.e.,
$^{204,202,200,198,196,194,192,190}_{\,~~~~~~~~~~~~~~~~~~~~~~~~~~~~~~~82}$Pb,
$^{124,122,120,118,116,112,108}_{\,~~~~~~~~~~~~~~~~~~~~~~~~~~~50}$Sn,
$^{64,62,60,58}_{\,~~~~~~~~~28}$Ni, and $^{50,48,46,44,42,40}_{\,~~~~~~~~~~~~~~~20}$Ca.
A theoretical error of $0.013$ MeV is assigned to $E_{B}$ and $0.011$ fm to
$r_c$ so that the respective ${\chi}^2$ is roughly equal to the number of the
corresponding data points. We further constrain the macroscopic parameters in
the optimization by requiring (1) the neutron $3p_{1/2}-3p_{3/2}$ energy level
splitting in $^{208}$Pb should lie in the range of $0.8-1.0$ MeV; (2) the
pressure of symmetric nuclear matter should be consistent with constraints
obtained from flow data in heavy ion collisions~\cite{Dan02}; (3) the binding
energy of pure neutron matter should be consistent with constraints obtained
from the latest chiral effective field theory calculations with controlled
uncertainties~\cite{Tew13}; (4) the critical density $\rho_{cr}$, above which
the nuclear matter becomes unstable by the stability conditions from Landau
parameters, should be greater than $2\rho_{0}$; and (5) $m_{s,0}^*$ should
be greater than $m_{v,0}^*$ and here we set $m_{s,0}^*-m_{v,0}^* =0.1m$
($m$ is nucleon mass in vacuum) to be consistent with the extraction from
global nucleon optical potentials constrained by world data on nucleon-nucleus
and (p,n) charge-exchange reactions~\cite{XuC10}. In the optimization, for a
fixed $E_{\text{sym}}({\rho _{c})}$ ($L({\rho}_{c})$), the contribution
of the $21$ data on $\Delta r_{np}$ of Sn isotopes (the $19$ $\Delta E$
data) is included in the weighted sum of ${\chi}^2$.

By using the optimized values for other macroscopic quantities at a fixed
$E_{\text{sym}}({\rho _{c}})$ ($L({\rho}_{c})$), we show the ${\chi}^2$
evaluated from the difference between theoretical and experimental
$\Delta E$ ($\Delta r_{np}$) as a function of $E_{\text{sym}}({\rho _{c}})$
($L({\rho}_{c})$) by solid circles in Fig.~\ref{chsqLcEsymc}.
From the new relation between  ${\chi}^2$ and $E_{\text{sym}}({\rho _{c}})$
($L({\rho}_{c})$), one can extract a value of
$E_{\text{sym}}({\rho _{c})}=26.65\pm0.20$ MeV and $L({\rho}_{c})=46.0\pm4.5$
MeV at $95\%$ confidence level. It is seen that the
$E_{\text{sym}}({\rho _{c})}=26.65\pm0.20$ MeV obtained with optimization is
larger than the $E_{\text{sym}}({\rho _{c}})=26.08\pm0.17$ MeV extracted using
MSL0 without optimization by about $0.6$ MeV, implying the correlations of
$\Delta E$ for heavy isotope pairs with $L(\rho_r)$, $G_{V}$, $G_{S}$,
$m_{s,0}^{\ast }$, and $W_{0}$ observed in Fig.~\ref{dEPb} play a certain
roles in the extraction of $E_{\text{sym}}({\rho _{c}})$. On the other hand,
it is remarkable to see that the extracted value of $L({\rho}_{c})=46.0\pm4.5$
MeV with optimization agrees well with $L({\rho}_{c})=47.3\pm4.5$ MeV obtained
using MSL0 without optimization, indicating the correlations of $\Delta r_{np}$
of heavy nuclei with other macroscopic quantities play minor roles in the
extraction of $L({\rho _{c}})$, as shown in Fig.~\ref{NskinPb208}.
Besides the $E_{B}$ and $r_c$ of the $25$ spherical even-even nuclei as well
as the constraints on the neutron $3p_{1/2}-3p_{3/2}$ energy level splitting
in $^{208}$Pb, the pressure of symmetric nuclear matter and the binding energy
of pure neutron matter, the critical density $\rho_{cr}$ and
$m_{s,0}^*-m_{v,0}^* =0.1m$, by further including simultaneously the $21$ data
on $\Delta r_{np}$ of Sn isotopes and the $19$ data of $\Delta E$ in the
optimization, we obtain a globally optimized parameter set of Skyrme force
denoted as MSL1 as shown in Table~\ref{MSL1}. Unlike the MSL0 interaction
which is obtained by directly using their empirical values for the macroscopic
parameters, the MSL1 interaction is obtained by fitting experimental data or
constraints by optimization. The main difference between MSL0 and MSL1 is the
latter predicts a significantly smaller (larger) value of $L(\rho_0)=45.25$
MeV ($G_{V}=68.74$ MeV$\cdot $fm$^{5}$) compared to the $L(\rho_0)=60$ MeV
($G_{V}=5$ MeV$\cdot $fm$^{5}$) from MSL0. As expected, the MSL1 nicely
reproduces the optimized values of $E_{\text{sym}}({\rho _{c}})$ and
$L({\rho}_{c})$ obtained from ${\chi}^2$ analysis with optimization for
$\Delta E$ and $\Delta r_{np}$, respectively, as shown by solid circles
in Fig.~\ref{chsqLcEsymc}.
\begin{table}[tbp]
\caption{Skyrme parameters in MSL1 (left side) and some
corresponding nuclear properties (right side).}
\label{MSL1}%
\begin{tabular}{lr||lr}
\hline\hline
Quantity & MSL1 & Quantity & MSL1 \\ \hline
$t_{0}$ (MeV$\cdot $fm$^{3}$) & $-1963.23$ & $\rho _{0}$ (fm$^{-3}$) & $0.1586$
\\
$t_{1}$ (MeV$\cdot $fm$^{5}$) & $379.845$ & $E_{0}$ (MeV) & $-15.998$ \\
$t_{2}$ (MeV$\cdot $fm$^{5}$) & $-394.554$ & $K_{0}$ (MeV) & $235.12$ \\
$t_{3}$ (MeV$\cdot $fm$^{3+3\sigma }$) & $12174.9$ & $m_{s,0}^{\ast }/m$ & $%
0.806$ \\
$x_{0}$ & $0.320770$ & $m_{v,0}^{\ast }/m$ & $0.706$ \\
$x_{1}$ & $0.344849$ & $E_{\text{sym}}(\rho_c)$ (MeV) & $26.67$ \\
$x_{2}$ & $-0.847304$ & $L(\rho_c)$ (MeV) & $46.19$ \\
$x_{3}$ & $0.321930$ & $G_{S}$ (MeV$\cdot $fm$^{5}$) & $126.69$ \\
$\sigma $ & $0.269359$ & $G_{V}$ (MeV$\cdot $fm$^{5}$) & $68.74$ \\
$W_{0}$ (MeV$\cdot $fm$^{5}$) & $113.62$ & $E_{\text{sym}}(\rho _{0})$ (MeV) & $%
32.33$ \\
 & &$L(\rho _{0})$ (MeV) & $45.25$ \\
\hline\hline
\end{tabular}%
\end{table}

\begin{figure}[tbp]
\includegraphics[scale=0.37]{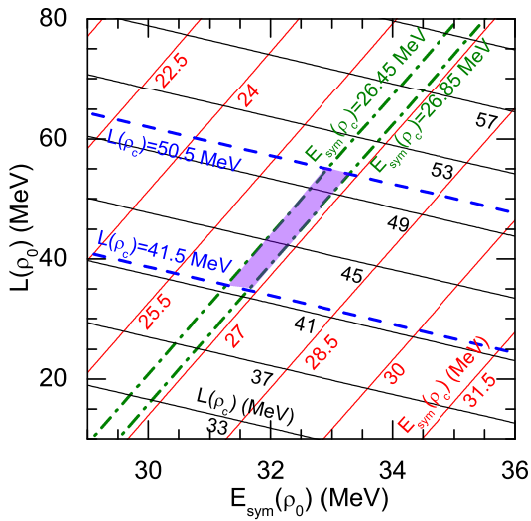}
\caption{(Color online) Contour curves in the $E_{\text{sym}}({\rho _{0}})$-$L(\rho_0)$
plane for $E_{\text{sym}}(\rho_c)$ and $L(\rho_c)$ from SHF calculation with MSL1.
The shaded region represents the overlap of constraints obtained from
$E_{\text{sym}}({\rho _{c})}=26.65\pm0.20$ MeV and $L({\rho}_{c})=46.0\pm4.5$ MeV.}
\label{LEsym}
\end{figure}
Since what we have directly constrained from $\Delta E$ and $\Delta r_{np}$ of Sn
isotopes are $E_{\text{sym}}({\rho _{c})}=26.65\pm0.20$ MeV and
$L({\rho}_{c})=46.0\pm4.5$ MeV at $\rho_c = 0.11$ fm$^{-3}$, the extrapolation
is necessary to obtain information on the symmetry energy at $\rho_0$ from
$E_{\text{sym}}({\rho _{c}})$ and $L({\rho}_{c})$. Shown in Fig.~\ref{LEsym} are
contours in the $E_{\text{sym}}({\rho _{0}})$-$L({\rho _{0}})$ plane for
$E_{\text{sym}}({\rho}_{c})$ and $L({\rho}_{c})$ from SHF calculation by
assuming the other $7$ macroscopic quantities $\rho _{0}$, $E_{0}(\rho_{0})$,
$K_{0}$, $m_{s,0}^{\ast }$, $m_{v,0}^{\ast }$, $G_{S}$, and $G_{V}$ are fixed
at their default (optimized) values in MSL1. It is interesting to see that a
fixed value of $L({\rho_c })$ leads to a negative
$E_{\text{sym}}({\rho _{0}})$-$L({\rho _{0}})$ correlation, which was
observed previously by directly constraining $E_{\text{sym}}({\rho _{0}})$
and $L({\rho _{0}})$ from analyzing the $\Delta r_{np}$ data of Sn
isotopes~\cite{Che10}. On the other hand, a fixed value of
$E_{\text{sym}}({\rho _{c})}$ leads to a positive
$E_{\text{sym}}({\rho _{0}})$-$L({\rho _{0}})$ correlation as expected
(See, e.g., Ref.~\cite{Che11}). Combing the constraints from
$E_{\text{sym}}({\rho _{c})}=26.65\pm0.20$ MeV and $L({\rho}_{c})=46.0\pm4.5$
MeV leads to a quite stringent constraint on both $E_{\text{sym}}({\rho _{0}})$
and $L({\rho _{0}})$ as indicated by the shaded region in Fig.~\ref{LEsym},
giving $E_{\text{sym}}({\rho _{0}})=32.3\pm1.0$ MeV and
$L({\rho _{0}})=45.2\pm10.0$ MeV, which is essentially consistent with
other constraints extracted from terrestrial experiments and astrophysical
observations as well as theoretical calculations with controlled
uncertainties~\cite{Tsa12,Lat12,ChenLW12,LiBA12,Tew13} but with much higher
precision. In particular, our results are in surprisingly good agreement
with the constraint of $E_{\text{sym}}({\rho _{0}}) = 31.2$-$34.3$ MeV and
$L({\rho _{0}}) = 36$-$55$ MeV (at 95\% confidence level) obtained recently
from Bayesian analysis of currently available neutron star mass and radius
measurements~\cite{Ste12} as well as that of
$E_{\text{sym}}({\rho _{0}}) = 29.0$-$32.7$ MeV and
$L({\rho _{0}}) = 40.5$-$61.9$ MeV obtained recently from the
experimental, theoretical and observational analyses~\cite{Lat12b}. Our
results also agree with the constraint of
$E_{\text{sym}}({\rho _{0}}) = 32.0\pm1.8$ MeV and
$L({\rho _{0}}) =  43.125\pm15$ MeV obtained by analyzing pygmy dipole
resonances (PDR) of $^{130,132}$Sn~\cite{Kli07} and that of
$E_{\text{sym}}({\rho _{0}}) = 32.3\pm1.3$ MeV and
$L({\rho _{0}}) =  64.8\pm15.7$ MeV from analyzing PDR of
$^{68}$Ni and $^{132}$Sn~\cite{Car10}. Furthermore, our results are
consistent with the constraint of $E_{\text{sym}}({\rho _{0}}) = 32.5\pm0.5$
MeV and $L({\rho _{0}}) =  70\pm15$ MeV obtained recently from a new and
more accurate finite-range droplet model analysis of the nuclear
mass of the 2003 Atomic Mass Evaluation~\cite{Mol12} although
the two constraints only have a small overlap for $L({\rho _{0}})$.

The core-crust transition density $\rho_t$ of neutron stars play crucial
roles in neutron star properties~\cite{Lat04} and it is strongly correlated
with the $L({\rho}_{0})$~\cite{XuJ09,Duc10}. We notice a similar strong
correlation also exists between $\rho_t$ and $L({\rho}_{c})$. Using a
dynamical approach (See, e.g., Ref.~\cite{XuJ09}), we obtain
$\rho_t=0.082 \pm 0.005$ fm$^{-3}$ from $L({\rho}_{c})=46.0\pm4.5$ MeV,
which agrees well with the empirical values~\cite{Lat04}. Furthermore,
we find the $L({\rho}_{c})=46.0\pm4.5$ MeV leads to a quite strong
constraint of $\Delta r_{np}=0.170 \pm 0.016$ fm for $^{208}$Pb, which
is in good agreement with the $\Delta r_{np}=0.156^{+0.025}_{-0.021}$ fm
obtained from the $^{208}$Pb dipole polarizability~\cite{Tam11} and
within the experimental error bar also consistent with the
$\Delta r_{np}=0.33^{+0.16}_{-0.18}$ fm extracted recently from the
PREX~\cite{Abr12}.

\section{Summary and outlook}

In summary, we show that while the binding energy difference $\Delta E$ between
a heavy isotope pair is essentially determined by the magnitude of the symmetry
energy $E_{\text{sym}}({\rho })$ at a subsaturation cross density
$\rho_c \approx 0.11$ fm$^{-3}$, the symmetry energy density slope $L({\rho })$
at the same $\rho_c$ fixes uniquely the neutron skin thickness $\Delta r_{np}$
of heavy nuclei. Our results demonstrate that the global properties (e.g., the
neutron skin thickness and binding energy difference) of heavy nuclei can be
effectively determined by the EOS of nuclear matter at a subsaturation cross
density $\rho_c \approx 0.11$ fm$^{-3}$ rather than at $\rho_0$, which is nicely
consistent with the recent finding in Ref.~\cite{Kha12} where the giant monopole
resonance of heavy nuclei has been shown to be constrained by the EOS of
symmetric nuclear matter at $\rho_c \approx 0.11$ fm$^{-3}$ rather than at
$\rho_0$.

Furthermore, we find a fixed value of $L({\rho_c })$ leads to a negative
correlation between $E_{\text{sym}}({\rho _{0}})$ and $L({\rho _{0}})$ at
saturation density $\rho_0$ while a fixed value of $E_{\text{sym}}({\rho _{c})}$
leads to a positive $E_{\text{sym}}({\rho _{0}})$-$L({\rho _{0}})$ correlation.
The existing data on $\Delta r_{np}$ of Sn isotopes and $\Delta E$ for a number
of heavy isotope pairs put simultaneously stringent constraints on the magnitude
and density slope of the $E_{\text{sym}}({\rho })$ at $\rho_c$, i.e.,
$E_{\text{sym}}({\rho _{c})}=26.65\pm0.20$ MeV and $L({\rho_c })= 46.0\pm4.5$
MeV at $95\%$ confidence level, whose extrapolation gives
$E_{\text{sym}}({\rho _{0}})=32.3\pm1.0$ MeV and $L({\rho _{0}})=45.2\pm10.0$
MeV. The obtained $E_{\text{sym}}({\rho _{0}})$ and $L({\rho _{0}})$ are
essentially consistent with other constraints extracted from
analyzing terrestrial experiments and astrophysical observations as well as
theoretical calculations with controlled uncertainties but with
higher precision. The extracted $L({\rho_c })$ also leads to
a strong constraint of $\Delta r_{np}=0.170 \pm 0.016$ fm for $^{208}$Pb and
$\rho_t=0.082 \pm 0.005$ fm$^{-3}$ for the core-crust transition density of
neutron stars.

Our results in the present work are only based on the standard SHF energy
density functional. It will be interesting to see how our results change if
different energy-density functionals, e.g., the extended non-standard SHF
energy density functional or relativistic mean field model, are used. These
studies are in progress and will be reported elsewhere.

\begin{acknowledgments}
This work was supported in part by the NNSF of China under Grant Nos. 10975097,
11135011, and 11275125, the Shanghai Rising-Star Program under grant No.
11QH1401100, the ``Shu Guang" project supported by Shanghai Municipal Education
Commission and Shanghai Education Development Foundation, the Program for Professor
of Special Appointment (Eastern Scholar) at Shanghai Institutions of Higher
Learning, and the Science and Technology Commission of Shanghai
Municipality (11DZ2260700).
\end{acknowledgments}

\end{document}